# Optimum Power and Rate Allocation for Coded V-BLAST: Average Optimization

Victoria Kostina, Sergey Loyka

*Abstract*—An analytical framework for performance analysis and optimization of coded V-BLAST is developed. Average power and/or rate allocations to minimize the outage probability as well as their robustness and dual problems are investigated. Compact, closed-form expressions for the optimum allocations and corresponding system performance are given. The uniform power allocation is shown to be near optimum in the low outage regime in combination with the optimum rate allocation. The average rate allocation provides the largest performance improvement (extra diversity gain), and the average power allocation offers a modest SNR gain limited by the number of transmit antennas but does not increase the diversity gain. The dual problems are shown to have the same solutions as the primal ones. All these allocation strategies are shown to be robust. The reported results also apply to coded multiuser detection and channel equalization systems relying on successive interference cancelation.

## I. Introduction

To exploit the impressive spectral efficiencies of wireless communication systems with multiple antennas at both the transmitter and receiver [1], the V-BLAST algorithm was proposed [2]. Its simple transmission and detection mechanisms as well as its ability to achieve a significant portion of the MIMO capacity have made the V-BLAST a popular solution for MIMO signal processing. In this paper, we consider zero-forcing (ZF) V-BLAST, which relies on the successive interference cancelation (SIC) to decode the spatially-multiplexed data streams at the receiver. Because of the SIC, the algorithm suffers from the error propagation effect so that the overall error performance is dominated by that of the 1st stream (with low diversity order) [3][4][5][6], which may not be satisfactory. While the optimal ordering procedure provides some improvement in SNR (the SNR gain of the ordering equals to the number of transmit antennas at high SNR in i.i.d. Rayleigh fading), the system diversity order is not improved and is still limited by that of the 1st stream (i.e. the lowest one) [4][7][8], which may be unsatisfactory.

Several techniques have been reported to improve the error performance of the uncoded V-BLAST by employing a non-uniform power allocation among the transmitters. In particular,

Manuscript received September 15, 2009, revised February 11, 2010, accepted September 10, 2010.

V. Kostina is with the Department of Electrical Engineering, Princeton University, NJ, 08544, USA, e-mail: vkostina@princeton.edu.

S. Loyka is with the School of Information Technology and Engineering, University of Ottawa, Ontario, Canada, K1N 6N5, e-mail: sergey.loyka@ieee.org.

[9][10][11] explore the transmit (Tx) power allocation that minimizes the instantaneous (i.e. for given channel realization) error rate of the uncoded V-BLAST, with or without the optimal ordering. For such optimization, a new feedback session and power reallocation are needed each time the channel changes due to small scale or multipath fading. A less demanding approach is to find an optimum allocation of the average power based on the average error rate, a strategy we refer to as "average optimization". Since this ignores small-scale fading, only occasional feedback sessions and power reallocations are required, when the average SNR changes due to large-scale fading or shadowing, and only the average SNR needs to be fed back to the transmitter. Average power allocation for the uncoded V-BLAST has been explored in [3][5][12]. While its performance is slightly inferior to the instantaneous power allocation, it does offer a few dB improvement in the SNR and achieves the same gain equal to the number of transmit antennas at high SNR [12]. The SNR gain of any optimum power allocation is upper bounded by the number of transmit antennas and therefore it does not improve the diversity gain (provided that all the streams stay active, which is required to support high rate) [12]. Another way to reduce error rate of the V-BLAST is via the fixed-complexity sphere decoder, which, under certain scenarios, demonstrates near-maximum likelihood performance with a fixed number of operations [13].

While the studies above deal with the uncoded V-BLAST, most practical communication systems use coding; uncoded systems are rare. A coded V-BLAST OFDM system has been considered in [14] via the system capacity analysis and an improved detection scheme to reduce the impact of error propagation has been presented, assuming uniform power allocation. An instantaneous optimization of power, rate and antenna mapping for a coded ZF V-BLAST to minimize the total transmit power for given data rate under a zero-outage constraint has been presented in [15], assuming capacity-achieving codes or realistic ones via the SNR gap to capacity. The optimization is performed by a numerical algorithm exploiting the problem convexity. While this approach allows significant improvement in system performance, it also requires instantaneous feedback and computations, which increases the system complexity. Due to the numerical nature of the solution, only limited insight is available. The SNR-asymptotic diversity-multiplexing tradeoff (DMT) analysis of the ordered V-BLAST with optimum rate allocation (to optimize the DMT) is presented in [16]. While this analysis provides some insight into the SNR asymptotics of the performance, it is not clear what the finite-SNR implications are.

In this paper, we use an analytical approach to optimization and performance analysis of coded ZF V-BLAST, which provides significant new insights at finite SNR. To reduce the demand on system resources, we consider average optimization of power and rate allocation and provide closed-form solutions. Following the earlier work in [14][15] and also the general philosophy advocated in [17][18], we assume that temporal capacity-achieving codes are used for each data stream of the V-BLAST, so that all per-stream rates up to the stream capacity can be supported with zero probability of error. This is motivated by the fact that there are powerful codes (LDPC, turbo-codes) that operate very close to capacity. Realistic codes are accounted for via the SNR gap to capacity, as in [15][19]. This model of coded V-BLAST allows analytically-tractable optimization and performance analysis of the algorithm.

We perform analysis and performance evaluation of the following three optimization strategies, which minimize the outage probability under the constraints on the total power and rate:

- average power allocation (APA), which is motivated by the fact that many practical system use power control,
- average rate allocation (ARA), which is suitable for variable-rate system using identical and fixed power amplifiers to simplify the RF part of the system,
- joint average power and rate allocation (APRA), which is suitable for variable-rate variable-power systems.

Compact, closed-form expressions for the optimized powers and rates are obtained and their error rate performance and robustness are investigated. Our approach is to use analytical techniques as much as possible and to resort to numerical techniques only for validation purposes.

The following novel insights are obtained:

- The APA does not change the system diversity order, but offers an SNR gain, which is upper bounded by the number of transmit antennas; the upper bound is achieved in the low outage regime. This is similar to the uncoded V-BLAST [12].
- The ARA improves the diversity order and thus is much more efficient than the APA in the low outage regime.
- While the APRA offers a performance improvement in high-to-moderate outage regime, it has the same performance as the ARA in the low outage one. Surprisingly, the uniform power allocation across the active streams is optimum in this regime.
- All these optimization strategies are robust in terms of rate and/or power variations (with power allocation demonstrating better robustness than rate allocation), which makes them good candidates for practical systems.
- Dual problems of minimizing the total power or maximizing the total rate under the outage probability constraint have the same solutions as the primal ones.

Comparing the 3 optimization strategies, we conclude that the ARA offers the largest incremental increase in the performance, with the APA offering a modest SNR gain over the unoptimized system and the APRA offering an improvement over the ARA in the high-to-moderate outage regime.

Due to similar system architectures and processing strategies, most of these results also apply to multiuser detection and inter-symbols interference equalization systems that use successive interference cancellation.

All analytical results and approximations are validated via simulations.

The paper is organized as follows. Section II introduces the basic system model, assumptions and optimization strategies. Sections III - V derive and analyze the optimum power, rate and joint power/rate allocations for the coded V-BLAST, section VI studies their robustness and section VII considers the dual problems. Section VIII concludes the paper.

## II. SYSTEM MODEL AND OUTAGE PROBABILITY

Analytical performance evaluation of the V-BLAST is a challenging task even when no coding is used [3][4][5][6], mainly because of the successive interference cancelation. Adding realistic codes to that would make the problem analytically intractable. On the other hand, many state-of-the art codes (e.g. turbo-codes, LDPC) operate very close to the capacity (within a fraction of a dB), so that assuming capacity-achieving codes is practically relevant and it also makes the analysis significantly simpler in many cases. This approach has been successfully exploited in [14][15][17][18] and will be applied here to the coded V-BLAST analysis and optimization.

In particular, we study ZF V-BLAST without optimal ordering but with optimized average power and/or rate allocation and capacity-achieving temporal codes for each stream so that the maximum possible rate equals to the capacity of that stream. Since capacity-achieving codes are used, there are no errors if the stream is not in outage and roughly half of the bits are in errors during an outage, so that the overall bit error rate is $\mathsf{BER} \approx \frac{1}{2} P_{out}$ [1], where $P_{out}$ is the outage probability, i.e. the probability that the system capacity is less than the target rate. There is no error propagation when all streams are not in outage, which also simplifies the analysis significantly.

The following standard baseband discrete-time MIMO system model is employed,

$$\mathbf{r} = \mathbf{H}\mathbf{\Lambda}\mathbf{s} + \boldsymbol{\xi} = \sum_{i=1}^{m} \mathbf{h}_i \sqrt{\alpha_i} s_i + \boldsymbol{\xi} \qquad (1)$$

where $\mathbf{s} = [s_1, s_2, ...s_m]^T$ and $\mathbf{r} = [r_1, r_2, ...r_n]^T$ are the vectors representing the Tx and Rx symbols respectively, "$T$" denotes transposition, $\mathbf{H} = [\mathbf{h}_1, \mathbf{h}_2, ...\mathbf{h}_m]$ is the $n \times m$ matrix of the complex channel gains between each Tx and each Rx antenna, where $\mathbf{h}_i$ denotes i-th column of $\mathbf{H}$, $n$ and $m$ are the numbers of Rx and Tx antennas respectively, $n \geq m$, $\boldsymbol{\xi}$ is the vector of circularly-symmetric additive white Gaussian noise (AWGN), which is independent and identically distributed (i.i.d.) in each receiver, $\mathbf{\Lambda} = \mathrm{diag}\left(\sqrt{\alpha_1}, \ldots, \sqrt{\alpha_m}\right)$, where $\alpha_i$ is the power allocated to the $i$-th transmitter. For the regular V-BLAST, the total power is distributed uniformly among the transmitters, $\alpha_1 = \alpha_2 = ... = \alpha_m = 1$. The channel is assumed to be non-ergodic ("slow block fading") and the key performance measure is the outage probability for given target bit rate [17]. Details of a mathematical model of the

---

[1] this approximation is due to the fact that the transition from low to high error rate regime is very sharp in capacity-approaching codes (see e.g. [20])

uncoded V-BLAST, on which our model of the coded V-BLAST is based, and its analysis can be found in [3][4][6][12] and are not repeated here. After the interference cancelation from already detected symbols and interference nulling from yet-to-be detected symbols, the equivalent scalar channel of i-th stream is

$$r'_i = \sqrt{\alpha_i}|\mathbf{h}_{i\perp}|s_i + \xi'_i \quad (2)$$

where $\mathbf{h}_{i\perp}$ is i-th column of the channel matrix projected onto the subspace perpendicular to that spanned by yet-to-be-detected symbols (i.e. by $[\mathbf{h}_{i+1}, \mathbf{h}_{i+2}, ...\mathbf{h}_m]$). This channel can support all the rates up to its instantaneous capacity

$$C_i = \ln(1 + \alpha_i|\mathbf{h}_{i\perp}|^2\gamma_0) \text{ [nat/s/Hz]}, \quad (3)$$

where $\gamma_0$ is the average SNR, so that the system outage probability is

$$P_{out} = 1 - \prod_{i=1}^{m}(1 - \Pr\{C_i < R_i\}); \quad (4)$$

where $R_i$ is the stream fixed target rate (which may be a function of the average SNR only), and the power allocation $\boldsymbol{\alpha} = [\alpha_1 \ldots \alpha_m]$ is the function of the average SNR $\gamma_0$ only. $\Pr\{C_i < R_i\}$ is the outage probability of i-th stream, and the system outage takes place when at least one of the streams is in outage, i.e. is not able to support its target rate. In i.i.d. Rayleigh fading channels, different $|\mathbf{h}_{i\perp}|^2$ are distributed as $\chi^2_{2(n-m+i)}$ (chi-squared with $2(n-m+i)$ degrees of freedom) and are independent of each other [3][6], so that the outage probability of $i$-th stream is equal to the outage probability of $(n-m+i)$-th order maximum ratio combiner (MRC):

$$P_i = \Pr\{C_i < R_i\} = \Pr\{\ln(1 + \alpha_i|\mathbf{h}_{i\perp}|^2\gamma_0) < R_i\}$$
$$= F_{n-m+i}\left(\frac{e^{R_i}-1}{\alpha_i\gamma_0}\right) \quad (5)$$
$$\approx \frac{1}{(n-m+i)!}\left(\frac{e^{R_i}-1}{\alpha_i\gamma_0}\right)^{n-m+i}, \quad \frac{e^{R_i}-1}{\alpha_i\gamma_0} \ll 1$$

where $F_k(x) = 1 - e^{-x}\sum_{l=0}^{k-1}x^l/l!$ is the outage probability of $k$-th order MRC. In the low outage regime, which is of practical interest, $R_i < \ln(1 + \alpha_i\gamma_0)$ and $P_i \ll 1$, so that (4) can be approximated by 1st order terms only

$$P_{out} \approx \sum_i P_i \approx \sum_i \frac{1}{(n-m+i)!}\left(\frac{e^{R_i}-1}{\alpha_i\gamma_0}\right)^{n-m+i} \quad (6)$$

This is the approximation we use throughout the paper (Lemma 6 in Appendix shows that an optimization preserves the approximation accuracy, which justifies this approach). We note that while the approximations similar to those in (5)(6) also hold for the uncoded system [3][4][6][12], there is a significant difference: the low outage regime implies high SNR ($\gamma_0 \gg 1$) for the uncoded system, but in the coded system the SNR may also be low in this regime as long as $R_i < \ln(1 + \alpha_i\gamma_0)$ (for example, it holds for multi-user interference-limited systems, where the interference is considered to be a part of noise, e.g. CDMA systems [17]). This observation also applies to the results below which are obtained via these approximations.

With realistic rather than capacity-achieving codes, the channel in (2) supports all the rates up to $\ln(1 + \alpha_i|\mathbf{h}_{i\perp}|^2\gamma_0/\Gamma)$, where $\Gamma$ is the SNR gap to capacity [19][15] so that all our results will also apply with the substitution $\gamma_0 \to \gamma_0/\Gamma$, i.e. with the "effective" SNR $\gamma_0/\Gamma$.

With a fixed power and rate allocation, including the unoptimized system ($R_i = R$, $\alpha_i = \alpha$), the diversity order of i-th stream is $n - m + i$ so that 1st stream has the lowest diversity order $n - m + 1$ and, thus, this stream provides the dominant contribution to the system outage probability in the low outage regime, with vanishingly small contribution coming from higher-order streams,

$$P_{out} \approx P_1 \gg P_2 \gg ... \gg P_m \quad (7)$$

and the overall system diversity order is also $n - m + 1$. This parallels the corresponding result for the uncoded system [3][6].

In this paper, we consider an average optimization, i.e. an optimum power and/or rate allocation is found based on channel statistics and stays the same as long as the average SNR stays the same so that the system tracks only large-scale channel variations (similarly to the uncoded system [3][12]), which reduces the demand on system resources and feedback channel.

The following optimization strategies are considered below: optimum average (per-stream) power allocation (APA), optimum average (per-stream) rate allocation (ARA), and joint average power and rate allocation (APRA), all to minimize the outage probability for given total data rate and power.

## III. AVERAGE POWER ALLOCATION

In this section, we consider the optimum average power allocation (APA) among the streams with uniform rate allocation, $R_i = R$:

$$\min_{\boldsymbol{\alpha}} P_{out} \quad \text{subject to} \quad \sum_{i=0}^{m}\alpha_i = m \quad (8)$$

The solution of this problem can be characterized in the following way.

**Theorem 1.** *In the low outage regime[2], $R < \ln(1 + \gamma_0)$, the optimum allocation of average power to minimize $P_{out}$ (i.e. the problem in (8)) can be approximated as*

$$\alpha_1^\star \approx m - \sum_{i=2}^{m}\alpha_i^\star, \quad (9)$$

$$\alpha_i^\star \approx b_i\left[\frac{e^R - 1}{\gamma_0}\right]^{\frac{i-1}{n-m+i+1}}, \quad i = 2\ldots m,$$

*where the numerical coefficients $b_i$ are given by*

$$b_i = \left[\frac{m^{n-m+2}(n-m)!}{(n-m+i-1)!}\right]^{\frac{1}{n-m+i+1}} \quad (10)$$

*Proof:* see Appendix A. ∎

---

[2]Note that $R < \ln(1 + \gamma_0)$ is the low outage regime for the unoptimized system. As the optimization can only decrease outage probability, the approximation in (5) is always valid for a system in which $\alpha_i$ and/or $R_i$ have been optimized in one way or another.

In the unoptimized system, 1st stream error rate has the smallest diversity order $n - m + 1$, which increases with the stream number (see (5)), so that the outage probability is dominated by 1st term at the low outage regime, $P_{out} \approx P_1$, and the APA allocates most of the power to this stream to reduce $P_{out}$, with progressively smaller amounts to higher-order streams, which can be formalized as follows.

**Corollary 1.** *The optimum power allocation in* (9) *behaves at the low outage regime as follows,*

$$\alpha_1^\star \approx m \gg \alpha_2^\star ... \gg \alpha_m^\star, \quad (11)$$

*i.e. most of the power goes to 1-st stream, with vanishingly small portions to higher-order streams.*[3]

The error rates of the optimized system are obtained by combining (5) and (9):

$$P_1^* \approx \frac{1}{(n-m+1)!} \left(\frac{e^R - 1}{m\gamma_0}\right)^{n-m+1}, \quad (12)$$

$$P_i^* \approx b_i^{-\frac{(i-1)(n-m+i)}{n-m+i+1}} \left(\frac{e^R - 1}{\gamma_0}\right)^{d_i},$$

where $d_i$ is i-th stream diversity order,

$$d_i = \frac{(n-m+i)(n-m+2)}{n-m+i+1} \quad (13)$$

Note that $d_{i+1} > d_i$ so that 1st stream dominates the outage with vanishingly smaller contribution from higher-order ones so that the following corollary holds.

**Corollary 2.** *The error rates with the APA behave in the same way as in the unoptimized system at the low outage regime,*

$$P_{out}^* \approx P_1^* \gg P_2^* \gg ... \gg P_m^* \quad (14)$$

To quantify the performance improvement of the APA, we use the SNR gain $G$ of optimum power allocation, which is defined as the difference in the SNR required to achieve the same error rate in the unoptimized and optimized systems [12],

$$P_{out}(\alpha_1^\star, ..., \alpha_m^\star) = P_{out}(G, ..., G) \quad (15)$$

**Lemma 1.** *The SNR gain $G$ of either average or instantaneous power allocation is bounded, for any fading distribution, as follows:*

$$1 \leq G \leq m \quad (16)$$

*In the i.i.d. Rayleigh fading channel, the upper bound is achieved at the low outage regime, when $R < \ln(1 + \gamma_0)$, via the APA in Theorem 1 or via an instantaneous power allocation that minimizes $P_{out}$.*

*Proof:* The lower bound follows from the fact that optimized system cannot perform worse than the unoptimized one. The upper bound is a gain of a hypothetical system in which *every* stream enjoys $m$-fold boost in power, $\alpha_i = m$. The power-constrained optimized system cannot perform better than that, since in this case the *total* available power equals $m$, so that $\alpha_i \leq m$, and $P_{out}$ is decreasing in each $\alpha_i$. The achievability part of the APA follow from the comparison of (12), (14) to (5), (6). Since instantaneous optimization cannot perform worse than average, this also proves the achievability part of the instantaneous optimization. ∎

---
[3]This is similar to the case of uncoded V-BLAST in [3][12].

The power allocation in (9) approaches the maximum SNR gain of $m$ in the low outage regime. Note that the bounds in (16) are the same as for the unoptimized V-BLAST in [12]: in either coded or uncoded V-BLAST system, power allocation cannot bring in a SNR gain greater than $m$. It thus follows that there is no additional diversity gain associated with the APA in both cases. The diversity gain can be found from [17]

$$d = -\lim_{\gamma_0 \to \infty} \ln P_{out} / \ln \gamma_0. \quad (17)$$

or, when a closed-form expression for $P_{out}$ is available (e.g. (12)), by inspection. The following corollary follows immediately from Lemma 1:

**Corollary 3.** *The average APA does not provide any additional diversity gain over the unoptimized system for any fading distribution. In the i.i.d. Rayleigh-fading channel,*

$$d_{APA} = n - m + 1 = d_u$$

*where $d_u = n - m + 1$ is the diversity gain of the unoptimized system and the target rate $R$ is fixed.*[4]

**Example.** To get some insight, consider the coded $2 \times 2$ V-BLAST. The unoptimized outage probabilities are given by

$$P_{out} \approx P_1 \approx \frac{e^R - 1}{\gamma_0} \gg P_2 \approx \frac{1}{2}\left[\frac{e^R - 1}{\gamma_0}\right]^2 \quad (18)$$

The stream diversity gains are $d_1 = 1 < d_2 = 2$. The optimum power allocation is given by

$$\alpha_1^\star \approx 2 - \alpha_2^\star \gg \alpha_2^\star \approx 4^{\frac{1}{3}}\left[\frac{e^R - 1}{\gamma_0}\right]^{\frac{1}{3}} \quad (19)$$

The resulting optimized outage probabilities are given by

$$P_{out}^* \approx P_1^\star \approx \frac{e^R - 1}{2\gamma_0} \gg P_2^\star \approx \frac{1}{2 \cdot 4^{\frac{2}{3}}}\left[\frac{e^R - 1}{\gamma_0}\right]^{\frac{4}{3}} \quad (20)$$

and the diversity gains are $d_1^* = 1 < d_2^* = 4/3$. The optimization reduces the outage probability of the dominating 1st stream by a factor of 2 so that $P_{out}^* \approx P_{out}/2$. However, the lower power allocated to the 2nd transmitter results in decreased diversity order at the 2nd stream, compared to the unoptimized system, $d_2^* < d_2 = 2$, yet $P_1^\star \gg P_2^\star$ so that $P_1^\star$ is still dominant in the low outage mode. The system outage probability for various fixed rates is plotted in Fig. 1. The approximation (9) of optimum $\boldsymbol{\alpha}^\star$ exhibits good accuracy over the whole SNR range. The SNR gain of APA increases with the SNR and approaches 3 dB (as predicted by (16)), which is the same as the SNR gain of the optimal ordering with the uniform power/rate allocation [4][8].

---
[4]Note that the diversity gain is the same as for the uncoded system [6][12]. For coded system (with capacity-achieving codes), this result has been obtained (in a different way) in [17][16].

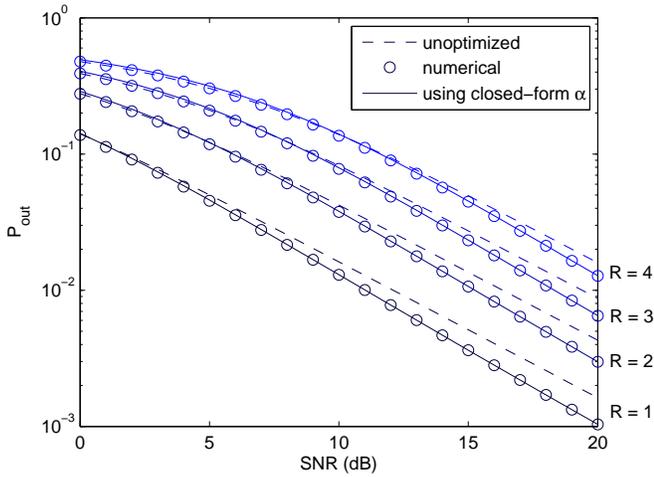

Fig. 1. The system outage probability of $2 \times 2$ V-BLAST in i.i.d. Rayleigh fading channel with and without power allocation for various fixed rates. The closed-form $\boldsymbol{\alpha}^\star$ in (9) provides good accuracy over the whole SNR range.

## IV. AVERAGE RATE ALLOCATION

In this section, we study the optimum average rate allocation (ARA) assuming the uniform power allocation, $\alpha_i = 1$. The optimization problem is formulated as follows:

$$\min_{R_1 \ldots R_m} P_{out} \quad (21)$$
$$\text{subject to } \sum_i R_i = mR, \ R_i \geq 0,$$

so that the total system target rate is $mR$ and individual rates are adjusted to minimize the system outage probability. The following Lemmas provide some insight into the problem and are also instrumental in finding the optimal rate allocation.

**Lemma 2.** *In the low outage regime $R < \ln(1 + \gamma_0)$, the optimization problem in (21) is convex and thus has a unique solution.*

*Proof:* see Appendix B. ∎

**Lemma 3.** *In the low outage regime, the optimum rates in the optimization problem above satisfy*

$$R_1^\star \leq R_2^\star \ldots \leq R_{m-1}^\star < R_m^\star \quad (22)$$

*Proof:* see Appendix B. ∎

**Lemma 4.** *In low outage/moderate rate regime $1 < R < \ln(1 + \gamma_0)$, the optimum per-stream outage probabilities at streams $i$ and $j$ are related as $(n-m+i)P_i^\star = (n-m+j)P_j^\star$ provided that both streams are active, so that $P_i^* > P_j^*$ for $i < j$, i.e. lower-order streams contribute more to the system outage probability.*

*Proof:* see Appendix B. ∎

Lemma 4 implies that the diversity orders at each stream are equal in the ARA, unlike the APA. Note the waterfilling analogy here: since the optimum is achieved when stream error rates are roughly equal, the rate allocation algorithm "pours" more rate to the streams with smaller initial error rate (higher diversity order streams) in order to make their error rate larger and less rate to lower order streams in order to make their error rate smaller. If some lower order streams are too weak, no rate is allocated to them.

We are now in a position to characterize the solution to the optimization problem in (21).

**Theorem 2.** *In low outage/moderate rate regime $1 < R < \ln(1 + \gamma_0)$, an approximate solution to the optimization problem in (21) is given by:*

$$R_i^\star \approx \left[ \ln \gamma_0 + \frac{mR - m_A \ln \gamma_0}{b(n-m+i)} + c_i \right]_+, \quad (23)$$

*where $[\cdot]_+ = \max[0, \cdot]$, and $a$, $b$ and $c_i$ are given by*

$$a = \sum_{i=1}^{m_A} \frac{\ln(n-m_A+i-1)!}{n-m_A+i}, \ b = \sum_{i=1}^{m_A} \frac{1}{n-m_A+i},$$
$$c_i = \frac{\ln(n-m+i-1)! - \frac{a}{b}}{n-m+i},$$

*and $m_A$ is the number of active transmitters, determined from*

$$m_A = \arg\min_{k=1:m} P_{out}(\boldsymbol{R}^*|_{m_A=k}) \quad (24)$$

*Proof:* see Appendix B. ∎

**Note.** Another way to find $m_A$ is based on Lemma 3:

$$m_A \approx \arg\max_{k=1:m} \left\{ k : R_{m-k+1}^\star|_{m_A=k} > 0 \right\} \quad (25)$$

While this is less accurate than (24) in some cases, it provides an additional insight unavailable from the latter.

Using (6), (23) and $P_i^\star = \frac{n}{n-m+i} P_m^\star$ (from Lemma 4), one obtains the rate-optimized outage probability

$$P_{out}^\star \approx n \sum_{i=m-m_A+1}^{m} \frac{1}{n-m+i} P_m^\star = nb P_m^\star \quad (26)$$
$$= \frac{b(n-m_A+m-1)!}{(n-1)!} e^{\frac{mR-a}{b}} \gamma_0^{-\frac{m_A}{b}},$$

where $b$ is as in (23).

It follows from Theorem 2 that the larger the target rate $mR$, the more transmitters stay active with the optimal rate allocation. If we begin to increase the SNR while keeping the total rate fixed, eventually only one stream will remain active at high SNR. While this approach, analyzed by Prasad and Varanasi in [3], achieves full receive diversity, it does not exploit the entire available capacity. On the other hand, if we set $mR = \overline{C} \approx m\ln(1+\gamma_0)$ to maximize the rate, where $\overline{C}$ is the ergodic system capacity, the outage probability becomes large, $P_{out} \approx \frac{1}{2}$. To balance the two extremes, we follow [17] and introduce the multiplexing gain $0 \leq r \leq m$, where $r = 0$ corresponds to a fixed rate at high SNR, and set $R = \frac{r}{m} \ln(1+\gamma_0) \approx \frac{r}{m} \ln \gamma_0$, which represents an adaptive transmission system where the data rate is directly linked to the available average SNR. Smaller $r$ decreases the outage probability, while larger $r$ permits higher transmission rates. Thus by adjusting the multiplexing gain, we control the rate-outage probability tradeoff. For this setting, the following corollary follows from Lemma 4 and (26).

**Corollary 4.** *In the low outage/moderate rate regime $1 < R = \frac{r}{m}\ln\gamma_0 < \ln(1+\gamma_0)$, the diversity gains at each stream are equal to the system diversity gain $d_{ARA}$:*

$$d_{ARA} = \left(\frac{1}{m_A}\sum_{i=0}^{m_A-1}\frac{1}{n-i}\right)^{-1}\left(1-\frac{r}{m_A}\right) \quad (27)$$

*which can be bounded as*

$$(n-m_A+1)\left(1-\frac{r}{m_A}\right) \leq d_{ARA} \leq n\left(1-\frac{r}{m_A}\right)$$

*where the upper bound is achieved when $m_A = 1$.*

When optimized over $m_A$, the DMT in (27) is the same as that in [16] for the fixed ordering with rate allocation. Note however that our result holds at finite SNR while that in [16] was derived for $SNR \to \infty$ (see [21] for a detailed discussion of limitations of the SNR-asymptotic DMT).

**Example.** Consider the $2 \times 2$ system, $n = m = 2$. With $R = \frac{r}{2}\ln(1+\gamma_0) \approx \frac{r}{2}\ln\gamma_0$, both transmitters are active when the rate is high:

$$r > \frac{1}{2} + \frac{3\ln 3}{4\ln\gamma_0} \approx \frac{1}{2} \quad (28)$$

where the inequality is from (24) and the equality is from (25). Clearly, the latter agrees with the former at high SNR and the critical multiplexing gain, above which both streams are active, is 1/2. Another way to use this condition is

$$\ln\gamma_0 > \frac{3\ln 3}{2(2r-1)}, \quad (29)$$

which gives an SNR threshold above which both streams are active for given $r$. Under these conditions, the optimal rates are given by

$$R_1^\star \approx R - \frac{1}{3}\Delta R, \ R_2^\star \approx R + \frac{1}{3}\Delta R, \quad (30)$$

where $\Delta R = \ln\gamma_0 - R$ represents the per-stream rate adjustment that minimizes the system outage probability. The optimized outage probabilities are

$$P_1^\star \approx \gamma_0^{-\frac{4}{3}(1-\frac{r}{2})}, \ P_2^\star \approx \frac{1}{2}P_1^*, \ P_{out}^\star \approx \frac{3}{2}P_1^* \quad (31)$$

and the diversity gains at both streams are equal: $d_{ARA}(r) = d_1^*(r) = d_2^*(r) = \frac{4}{3}(1-r/2) > d_{APA}(r) = 1 - r/2$. This differs drastically from the APA in (20), where the 1st stream diversity gain does not improve compared to the unoptimized system and the 2nd stream error rate is asymptotically negligible. For the ARA, 2nd stream outage probability provides a sizable contribution to the system one and the system diversity gain is improved by the optimization.

The first transmitter is inactive when the opposite inequality holds in (28). In this "low-rate" regime, $R_2^\star = 2R$, and the optimized outage probability is given by

$$P_{out}^\star = P_2^\star \approx \frac{1}{2}\gamma_0^{-2(1-r)} \quad (32)$$

so that the diversity gain $d_{ARA}(r) = 2(1-r) > d_{APA}(r) = 1 - r/2$, i.e. a significant improvement over the APA. For fixed $R$, $d_{ARA} = 2 > d_{APA} = 1$, i.e. the ARA enjoys the full diversity as opposed to the APA.

The optimum outage probability and rate allocation for this example are plotted in Fig. 2. Note that the approximations in Theorem 2 agree well with the accurate numerical solutions over the whole SNR range.

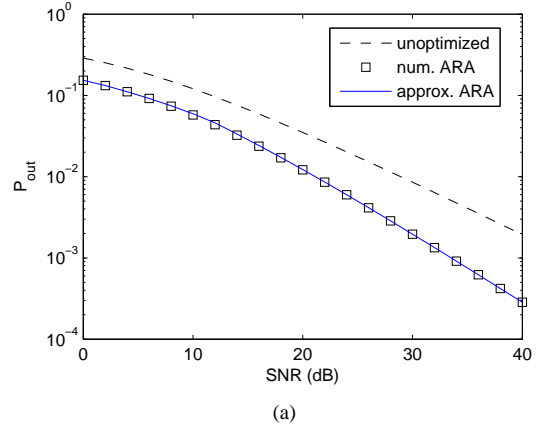

(a)

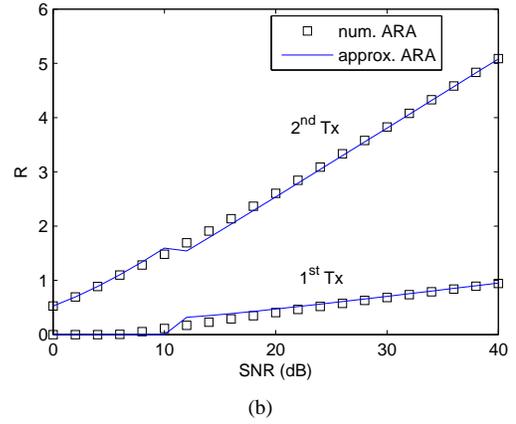

(b)

Fig. 2. Outage probability (a) and optimum rates (b) for $2 \times 2$ V-BLAST with the ARA in the i.i.d. Rayleigh fading channel, closed-form $R_i^\star$ is from (23), $r = 2/3$. Note that the approximation agrees well with the precise numerical solution, and that the ARA offers a significant reduction in the outage probability, even at low to moderate SNR.

## V. JOINT AVERAGE POWER/RATE ALLOCATION (APRA)

In this section, we consider the joint average power and rate allocation:

$$\min_{\boldsymbol{\alpha},\mathbf{R}} P_{out}, \ \text{subject to} \ \sum_i R_i = mR, \ R_i \geq 0, \quad (33)$$

$$\sum_i \alpha_i = m, \ \alpha_i \geq 0,$$

where the total rate is $mR$ and the total power is $m$, both of them are functions of the average SNR. The following property is instrumental.

**Lemma 5.** *In the low-outage moderate-rate regime, i.e. when $\ln 2 < R_i < \ln(1+\gamma_0)$ for all active streams, the optimization problem above is convex.*

*Proof:* see Appendix C. ∎

The solution can now be characterized in a somewhat surprising way.

**Theorem 3.** *In the low-outage moderate-rate regime $1 < R_i < \ln(1+\gamma_0)$, the uniform power allocation among active transmitters attains the minimum outage probability in (33),*

$$\alpha_i^\star = 0, \ i = 1, \ldots, m - m_A$$
$$\alpha_i^\star = \frac{m}{m_A}, \ i = m - m_A + 1, \ldots, m \qquad (34)$$

*The optimal rates and the number of active transmitters are given by (23), (24), respectively, with the substitution $\gamma_0 \to \frac{m}{m_A}\gamma_0$.*

*Proof:* see Appendix C. ∎

The APRA outage probability is given by (26) with $\gamma_0$ replaced by $\frac{m}{m_A}\gamma_0$, so that the APRA gains $\frac{m}{m_A}$ in the SNR compared to the ARA.

According to Theorem 3, a non-uniform power allocation on top of the rate allocation does not bring in any additional advantage in terms of the outage probability. The difference between the ARA and APRA lies in the extra power which is added to active transmitters in the APRA every time one of the transmitters is turned off while the per-transmitter power in the ARA is kept fixed. A consequence of this is that weak streams are kept silent by the APRA for a wider range of the SNR compared to the ARA (i.e. threshold multiplexing gain is higher for the former).

**Example.** Consider again the $2 \times 2$ system. As long as both transmitters are active, there is no difference with the ARA, and the optimal rates are given by (30), and the outage probabilities are as in (31). However, the threshold multiplexing gain or SNR are now slightly higher:

$$r > \frac{1}{2} + \frac{3\ln 12}{4\ln\gamma_0} \ \ \text{or} \ \ \ln\gamma_0 > \frac{3\ln 12}{2(2r-1)} \qquad (35)$$

This is a consequence of an extra redistributed power when one stream is turned off.

When only one transmitter is active, i.e. with the opposite inequality in (35), the outage probability becomes

$$P_{out}^\star = P_2^\star \approx \frac{1}{8}\gamma_0^{-2(1-r)} \qquad (36)$$

i.e. exhibits 4-fold improvement compared to the ARA in (32) with one active transmitter (regardless of the multiplexing gain), due to the doubled power on the remaining transmitter in the APRA. Note also that the APRA brings in additional diversity gain compared to the ARA in the range

$$\frac{1}{2} + \frac{3\ln 3}{4\ln\gamma_0} < r < \frac{1}{2} + \frac{3\ln 12}{4\ln\gamma_0}$$

because of the higher threshold.

The optimum outage probability and powers/rates are shown in Fig. 3 and 4 for various optimization strategies for $r = 1$. While the ARA and the APRA are identical for about $\gamma_0 > 15$ dB, when both streams are active in either system, the APRA exhibits better performance at lower SNR $\gamma_0 < 15$ dB, when there is only one active stream in the latter. Note also the difference in optimal power allocation between the APA and APRA strategies: while a larger portion of power goes to the 1st transmitter in the former, the power is distributed uniformly among active transmitters in the latter. The reason for this difference is that the stream diversity orders are equal in the rate-optimized system so there is no need to favor the first stream in order to reduce its dominating error rate. Overall, the APRA exhibits the best performance, the ARA approaches it at high SNR, followed by the APA, which still provides about 3 dB performance improvement over the unoptimized system at high SNR, which is the same as the SNR gain due to the optimal ordering in the unoptimized system [4][8].

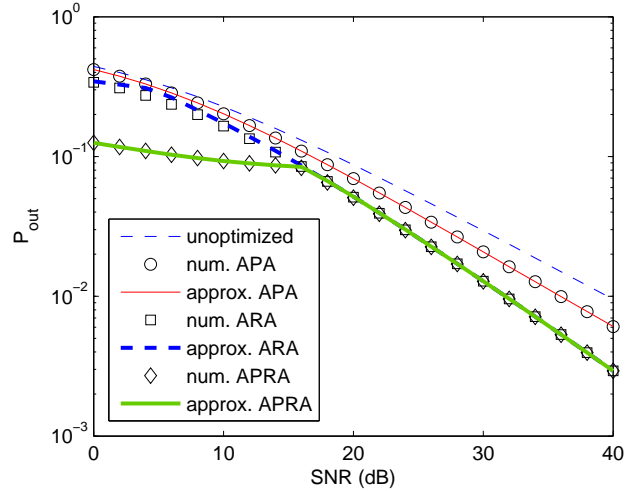

Fig. 3. The optimized outage probability of $2 \times 2$ V-BLAST in i.i.d. Rayleigh fading, $r = 1$, for various optimization strategies; the optimum power/rate allocations are as in Fig. 4.

## VI. ROBUSTNESS

Due to uncertainties or variations in system parameters and in the objective and constraint functions, a robust algorithm, which is insensitive to these variations, is desired from the practical perspective. In this section we discuss the robustness of the APA, ARA and APRA optimization strategies.

To quantify the robustness of the optimization algorithms above, we employ the following measure of local sensitivity to a system parameter $u$, which may be power or rate:

$$\delta = \left|\frac{\Delta P_{out}^\star / P_{out}^\star}{\Delta u / u}\right| \approx \left|\frac{dP_{out}^\star}{du}\frac{u}{P_{out}^\star}\right| \qquad (37)$$

where $P_{out}^\star$ denotes the outage probability of the optimized system so that $\delta$ is the ratio of the normalized variation in performance to the normalized variation in a system parameter. If $\delta$ is small to moderate number, it implies that a small change in $u$ leads to a relatively small change in $P_{out}^\star$. We then say that the algorithm is robust to variations in $u$. We are interested in how a small variation in $u = R_i^\star$ or $u = \alpha_i^\star$ affects the outage probability. From the optimality condition, $\partial P_{out}^\star / \partial u^\star = \nu$, where $\nu$ is the Lagrange multiplier in the corresponding convex optimization problem, so that

$$\delta = \nu u^\star / P_{out}^\star \qquad (38)$$

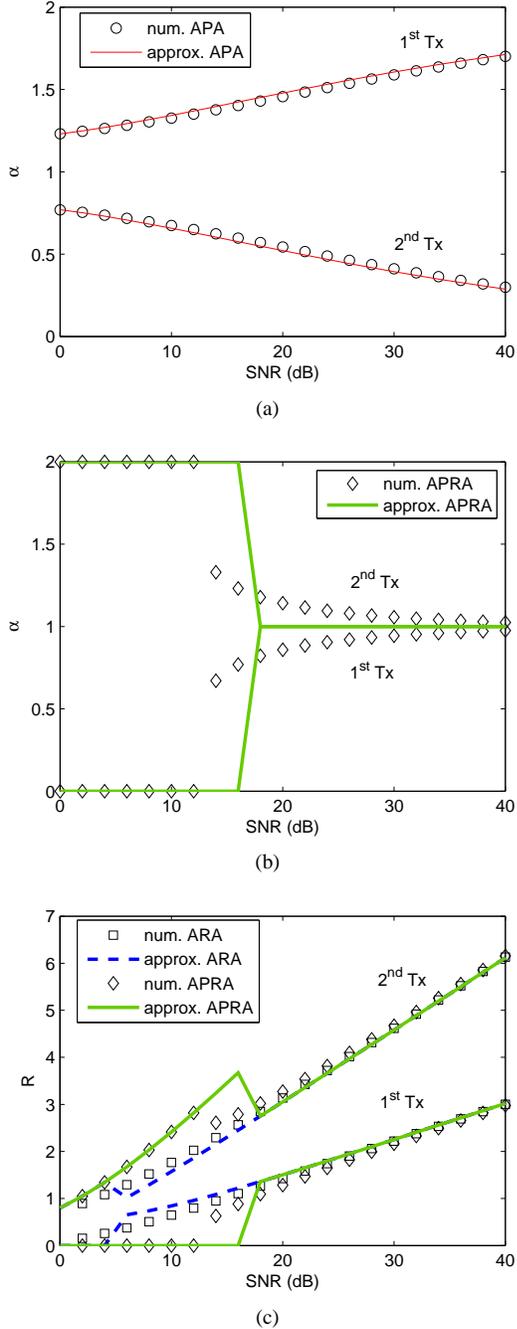

Fig. 4. The optimum powers for the APA (a) and for the APRA (b), and the rates (c), for the same setting as in Fig. 3. The approximate $\alpha_i^\star$ are from (9) for the APA and from (34) for the APRA; the approximate $R_i^\star$ are from (23). Note that the approximations agree well with the precise numerical solutions over the whole SNR range and that the ARA and APRA indeed have the same performance in the low outage regime. Also note the opposite tendencies in power allocations of the APA and APRA in (a) and (b): while the former allocates vastly different powers, the latter is almost uniform at high SNR. The SNR threshold in (35) is $\gamma_0 > 16$ dB and agrees well with Fig. 3(b).

i.e. the system robustness measure is proportional to the Lagrange multiplier [22].

Analysis of the APA robustness, $u = \alpha_i^\star$, is done similarly to the uncoded systems in [12], and (38) reduces to

$$\delta_1 \approx n - m + 1, \tag{39}$$

$$\delta_i \approx b_i \frac{n-m+1}{m} \left[\frac{e^R - 1}{\gamma_0}\right]^{\frac{i-1}{n-m+i+1}}, \; i \geq 2,$$

where $\delta_i$ denotes sensitivity to $\alpha_i$, and $b_i$ are given by (10). We see that the APA is robust as long as $n - m$ is not too large; moreover, since $\delta_1 \gg \delta_2 \gg \ldots \gg \delta_m$, higher streams exhibit better robustness.

For the ARA, we are interested in how a small variation in $R_i^\star$ affects the outage probability, $u = R_i^\star$. Substituting (53) and (26) into (38), one obtains:

$$\delta_{R_i} \approx \frac{(n-1)!}{b(n-m_A+m-1)!} R_i^\star < \frac{nm}{m_A} R, \tag{40}$$

where $mR$ is the total target rate, and the inequality is based on the observation that $n - 1 \leq n + m - m_A - 1$ and $b = \sum_{i=1}^{m_A} \frac{1}{n-m_A+i} \geq \frac{m_A}{n}$. When $R$ is proportional to $\ln \gamma_0$, which is required to obtain a non-zero multiplexing gain, and $mn/m_A$ is not too large, $\delta_{R_i}$ is a moderate number in the practical SNR range. Contrary to the APA case, lower stream demonstrate better robustness in the ARA, since $R_i^\star \leq R_{i+1}^\star$ (Lemma 3). In the case of $n = m = m_A$, (40) simplifies to

$$\delta_{R_i} \approx R_i^\star / b < R_i^\star, \tag{41}$$

where $b = \sum_{i=1}^m \frac{1}{i} > 1$.

For the APRA, it is easy to verify that the rate sensitivity is given by (40), while the power sensitivity is given by

$$\delta_\alpha \approx \frac{(n-1)!}{b(n-m_A+m-1)!} < \frac{n}{m_A}, \tag{42}$$

which says that the APRA is robust as long as $\frac{n}{m_A}$ is not too large. In contrast the APA, each stream here has exactly the same sensitivity to variations in its power.

**Example.** For $n = m = 2$, the APA sensitivities to first and second stream powers are given by

$$\delta_1 \approx 1 \gg \delta_2 \approx \left(2(e^R - 1)/\gamma_0\right)^{1/3}. \tag{43}$$

For the APRA with when both transmitters active, the sensitivities to $\alpha_i^\star$, $i = 1, 2$, are the same and equal to $2/3$, and the sensitivities to $R_1^\star$, $R_2^\star$ are given by

$$\delta_{R_1} \approx \frac{2}{3}\left(R - \frac{1}{3}\Delta R\right), \; \delta_{R_2} \approx \frac{2}{3}\left(R + \frac{1}{3}\Delta R\right), \tag{44}$$

where $\Delta R = \ln \gamma_0 - R$. When the first transmitter is turned off, the power sensitivity is 1 and the rate sensitivity is $\delta_2 \approx R$.

## VII. DUAL PROBLEMS

We have considered above the optimal power and/or rate allocations to minimize the outage probability. Due to the convex nature of these problems (which also have zero duality gap), these allocations can also be used to minimize the total power or maximize the total rate under the outage probability constraint - another practically-important possibility. This is formalized below.

**Theorem 4.** *Consider the following problem dual to* (8):

$$\min_{\boldsymbol{\alpha}} \sum_{i=0}^{m} \alpha_i \quad \text{subject to} \quad P_{out}(\boldsymbol{\alpha}) \leq \epsilon \quad (45)$$

*Its solution $\boldsymbol{\alpha^0}$ is the same as that of the problem in* (8) *and, in particular, is as in* (9) *under the conditions of Theorem 1 for $\epsilon = P_{out}(\boldsymbol{\alpha^*})$.*

*Proof:* see Appendix D ∎

Similar duality property hold for the problem in (21), which is formalized below.

**Theorem 5.** *Consider the following problem dual to* (21):

$$\max_{\boldsymbol{R}} \sum_{i=0}^{m} R_i, \quad \text{subject to} \quad P_{out}(\boldsymbol{R}) \leq \epsilon \quad (46)$$

*Its solution $\boldsymbol{R^0}$ is the same as that of the problem in* (21) *so that Theorem 2 applies with $\epsilon = P_{out}(\boldsymbol{R^*})$.*

*Proof:* along the same lines as that of Theorem 4. ∎

## VIII. CONCLUSION

A comparative analysis of the optimum power, rate and joint power and rate allocations for the coded V-BLAST has been presented. All considered optimization strategies use the average channel statistics rather than instantaneous channel matrix. This has the advantage of lower complexity due to less frequent feedback sessions and lighter computation load yet offers a significant improvement in performance. Compact, closed-form expressions for the optimum allocations of average power and rate have been given. The average rate allocation is the most rewarding strategy in terms of incremental improvement, since it improves the system diversity gain. In contrast, power allocation can at most give an $m$-fold SNR gain and it does not provide any additional diversity (provided that all streams stay active, as required for high spectral efficiency at high SNR). In the low outage regime, the uniform power allocation is already optimal for the coded V-BLAST with rate allocation, and using non-uniform power allocation does not improve the outage probability. This is very different from the power allocation only, where most of the power goes to the 1st stream with progressively smaller fractions to higher-order ones. All these optimization strategies are shown to be robust that makes them good candidates for practical systems. Dual problems of minimizing the total power or maximizing the total rate under the outage probability constraint are shown to have the same solutions as the primal ones.

## REFERENCES

[1] G. Foschini and M. Gans, "On limits of wireless communications in a fading environment when using multiple antennas," *Wireless personal communications*, vol. 6, no. 3, pp. 311–335, 1998.
[2] G. Foschini, G. Golden, R. Valenzuela, and P. Wolniansky, "Simplified processing for high spectral efficiency wireless communication employing multi-element arrays," *IEEE Journal on Selected Areas in Communications*, vol. 17, no. 11, pp. 1841–1852, 1999.
[3] N. Prasad and M. Varanasi, "Analysis of decision feedback detection for MIMO Rayleigh-fading channels and the optimization of power and rate allocations," *IEEE Transactions on Information Theory*, vol. 50, no. 6, pp. 1009–1025, 2004.
[4] S. Loyka and F. Gagnon, "Performance analysis of the V-BLAST algorithm: an analytical approach," *IEEE Transactions on Wireless Communications*, vol. 3, no. 4, pp. 1326–1337, July 2004.
[5] J. Choi, "Nulling and Cancellation Detector for MIMO Channels and its Application to Multistage Receiver for Coded Signals: Performance and Optimization," *IEEE Transactions on Wireless Communications*, vol. 5, no. 5, pp. 1207–1216, May 2006.
[6] S. Loyka and F. Gagnon, "V-BLAST without optimal ordering: analytical performance evaluation for Rayleigh fading channels," *IEEE Transactions on Communications*, vol. 54, no. 6, pp. 1109–1120, June 2006.
[7] Y. Jiang, X. Zheng, and J. Li, "Asymptotic Performance Analysis of V-BLAST," *IEEE GlobeCom, St. Louis, MO*, pp. 3882–3886, Nov. 28-Dec. 2, 2005.
[8] S. Loyka and F. Gagnon, "On Outage and Error Rate Analysis of the Ordered V-BLAST," *IEEE Transactions on Wireless Communications*, vol. 7, no. 10, pp. 3679–3685, Oct. 2008.
[9] S. Nam, O. Shin, and K. Lee, "Transmit power allocation for a modified V-BLAST system," *IEEE Transactions on Communications*, vol. 52, no. 7, pp. 1074–1079, 2004.
[10] R. Kalbasi, D. Falconer, and A. Banihashemi, "Optimum power allocation for a V-BLAST system with two antennas at the transmitter," *IEEE Communications Letters*, vol. 9, no. 9, pp. 826–828, 2005.
[11] N. Wang and S. Blostein, "Approximate minimum BER power allocation for MIMO spatial multiplexing systems," *IEEE Transactions on Communications*, vol. 55, no. 1, pp. 180–187, 2007.
[12] V. Kostina and S. Loyka, "On optimum power allocation for the V-BLAST," *IEEE Transactions on Communications*, vol. 56, no. 6, pp. 999–1012, June 2008.
[13] L. Barbero and J. Thompson, "Fixing the Complexity of the Sphere Decoder for MIMO Detection," *IEEE Transactions on Wireless Communications*, vol. 7, no. 6, pp. 2131–2142, June 2008.
[14] H. Lee and I. Lee, "New Approach to Error Compensation in Coded V-BLAST OFDM Systems," *IEEE Transactions on Communications*, vol. 55, no. 2, pp. 345–355, 2007.
[15] R. Zhang and J. Cioffi, "Approaching MIMO-OFDM capacity with zero-forcing V-BLAST decoding and optimized power, rate, and antenna-mapping feedback," *IEEE Transactions on Signal Processing*, vol. 56, no. 10, pp. 5191–5203, Oct. 2008.
[16] Y. Jiang and M. Varanasi, "Spatial Multiplexing Architectures with Jointly Designed Rate-Tailoring and Ordered BLAST Decoding - Part I: Diversity-Multiplexing Tradeoff Analysis," *IEEE Transactions on Wireless Communications*, vol. 7, no. 8, pp. 3252–3261, August 2008.
[17] D. Tse and P. Viswanath, *Fundamentals of wireless communication*. Cambridge Univ Press, 2005.
[18] G. Caire and K. Kumar, "Information Theoretic Foundations of Adaptive Coded Modulation," *Proceedings of the IEEE*, vol. 95, no. 12, pp. 2274–2298, 2007.
[19] J. Cioffi, *Digital Communications (course notes)*. Stanford University, 2007.
[20] J. Tillich and G. Zemor, "The Gaussian Isoperimetric Inequality and Decoding Error Probabilities for the Gaussian Channel," *IEEE Transactions on Information Theory*, vol. 50, no. 2, pp. 328–331, Feb. 2004.
[21] S. Loyka and G. Levin, "Finite-SNR Diversity-Multiplexing Tradeoff via Asymptotic Analysis of Large MIMO Systems," *IEEE Transactions on Information Theory*, vol. 56, no. 10, pp. 4781–4792, Oct. 2010.
[22] L. Boyd, S. Vandenberghe, *Convex Optimization*. Cambridge Univ Press, 2004.

## APPENDIX

### A. Proof of Theorem 1

It is straightforward to verify that this is a strictly convex problem and thus has a unique solution; the KKT conditions are necessary and sufficient for optimality. Taking the derivative with respect to $\alpha_i$ of the Lagrangian $L(\alpha) = P_{out}(\boldsymbol{\alpha}) + \lambda \left( \sum_i^m \alpha_i - m \right)$, where the approximation in (6) of $P_{out}$ is used, and equating it to zero according to the method of Lagrange multipliers, we obtain

$$\frac{\partial L(\alpha^\star)}{\partial \alpha_i^\star} = -\frac{\left(e^R - 1\right)^{n-m+i}}{\alpha_i^{\star n-m+i+1}\gamma_0^{n-m+i}(n-m+i-1)!} + \lambda = 0. \quad (47)$$

Observe that all of the $\alpha^\star_i$ can be expressed via the single parameter $\lambda$. Expressing $\alpha_i^\star$ via $\lambda$ and substituting the result into the total power constraint $\sum_i^m \alpha_i = m$, one obtains a polynomial equation for $\lambda$, which can be approximately solved using the Newton-Raphson method. The solution follows along the same lines as that for the uncoded system in Appendix A of [12] where further details of this method can be found. ∎

The following Lemma shows that an optimization preserves the accuracy of an approximation of the objective function, which justifies using (6) instead of the true outage probability.

**Lemma 6.** *Let $f(x)$ and $\hat{f}(x)$ be true and approximate objective functions, within $\pm \epsilon \%$ of each other,*

$$1 - \epsilon \leq \frac{\hat{f}(x)}{f(x)} \leq 1 + \epsilon$$

*Then, their optimized values are also within $\pm \epsilon \%$ of each other,*

$$1 - \epsilon \leq \frac{\min_x \hat{f}(x)}{\min_x f(x)} \leq 1 + \epsilon \quad (48)$$

*i.e. the original accuracy is preserved by optimization* [5].

*Proof:* Let $x^*$ and $\hat{x}^*$ be the true and approximate optimal points,

$$x^* = \arg\min_x f(x), \quad \hat{x}^* = \arg\min_x \hat{f}(x)$$

Note that $x^* \neq \hat{x}^*$ in general. Observe however that $\hat{f}(\hat{x}^*)/f(x^*) \leq 1 + \epsilon$. Indeed, assuming that the contrary is true, one obtains

$$\frac{\hat{f}(\hat{x}^*)}{f(x^*)} > 1 + \epsilon \geq \frac{\hat{f}(x^*)}{f(x^*)}$$

from which is follows that $\hat{f}(x^*) < \hat{f}(\hat{x}^*)$, which is impossible. Similar argument proves $\hat{f}(\hat{x}^*)/f(x^*) \geq 1 - \epsilon$ and thus (48). ∎

### B. Proof of Lemmas 2, 3, 4 and Theorem 2

The Lagrangian for this problem is

$$L = \sum_i^m P_i - \nu \left(\sum_i^m R_i - mR\right) - \sum_i^m \lambda_i R_i$$

Approximating $P_{out}$ according to (6), one obtains the KKT conditions as follows:

$$R_i^\star \geq 0, \; \sum_i R_i^\star = mR, \; \lambda_i \geq 0, \; \lambda_i R_i^\star = 0$$
$$P_i'(R_i^\star) - \lambda_i - \nu = 0, \; i = 1\ldots m \quad (49)$$

---

[5] The authors have originally proved a weaker version of this Lemma. It was then noted by E. Telatar that the present version may be true.

where $P_i'(R_i)$ denotes the derivative of $P_i$ with respect to $R_i$,

$$P_i'(R_i) \approx \frac{e^{R_i}\left(e^{R_i} - 1\right)^{n-m+i-1}}{\gamma_0^{n-m+i}(n-m+i-1)!}. \quad (50)$$

*Proof of Lemma 2 (convexity):* Notice that the second derivative of $P_i$ with respect to $R_i$ is positive for $R_i > 0$,

$$P_i''(R_i) = \frac{e^{R_i}\left(e^{R_i} - 1\right)^{n-m+i-2}\left[(n-m+i)e^{R_i} - 1\right]}{\gamma_0^{n-m+i}(n-m+i-1)!} > 0.$$

It follows that $P_{out}$ is convex as a sum of convex functions, and since the constrains are also convex, the problem in (21) is convex. Therefore the optimum rate allocation is unique in the low outage regime. ∎

Note that in the KKT conditions (49), $\lambda_i$ is a slack variable, which allows us to rewrite (49) as:

$$\begin{cases} P_i'(R_i^\star) - \nu \geq 0 \\ \left[P_i'(R_i^\star) - \nu\right] R_i^\star = 0 \end{cases} \quad (51)$$

*Proof of Lemma 3 (optimal rates are ordered):* Provided that all $R_i^\star$ are strictly positive, the optimum rate $\mathbf{R}^\star$ is the point at which the slope $P_i'$ is the same for all $i$, $P_i'(R_i^\star) = \nu$, as seen from from the KKT conditions (51). Let us fix $\gamma_0$ and $R_i = R$ and consider $P_i'$ as a function of $i$. Observe that $P_{i+1}'(R) = P_i'(R)\frac{e^R - 1}{\gamma_0(n-m+i)} < P_i'(R)$ as long as $e^R - 1 < \gamma_0$, which is satisfied in the low outage regime. In other words, $P_i'$ is monotonically decreasing in $i$. On the other hand, recall that $P_i'(R)$ as function of $R$ is monotonically increasing. We conclude that to achieve the equality $P_{i+1}'(R_{i+1}) = P_i'(R_i)$, it is necessary that $R_{i+1} > R_i$ for positive $R_i$.

Suppose now that for some $i$, $R_i^\star > 0$ but $R_{i+1}^\star = 0$. According to the KKT conditions (51), in this case $P_{i+1}'(R_{i+1}^\star) \geq \nu = P_i'(R_i^\star)$. But from the monotonicity properties of $P_i'$ discussed above, $P_{i+1}'(R_{i+1}^\star = 0) \leq P_i'(0) < P_i'(R_i^\star > 0)$ - a contradiction. We conclude that $R_1 \leq R_2 \ldots \leq R_{m-1} < R_m$ and it holds with equality between $R_i$ and $R_{i+1}$ if and only if $R_i = R_{i+1} = 0$. ∎

*Proof of Lemma 4 (diversity order):* it is easy to verify from (5), (50) that $P_i' = \frac{(n-m+i)e^{R_i}}{e^{R_i}-1}P_i \approx (n-m+i)P_i$. But at the optimum in (51), the slope $P_i'$ is the same for all active transmitters, $\nu = P_i'^\star \approx (n-m+i)P_i^\star$, $R_i^\star > 0$, so that the optimal stream error rates $P_i^\star$ are scaled versions of one another. ∎

*Proof of Theorem 2 (closed form rates):* From the KKT conditions in (51) we see that if $R_i^\star > 0$, then $P_i'(R_i^\star) = \nu$, or

$$e^{R_i}\left(e^{R_i} - 1\right)^{n-m+i-1} = \nu\gamma_0^{n-m+i}(n-m+i-1)!,$$

Let us employ the moderate to high rate approximation in the LHS, $e^{R_i}\left(e^{R_i} - 1\right)^{n-m+i-1} \approx e^{(n-m+i)R_i}$ (which becomes equality for $n - m + i = 1$), and take the logarithm of both sides:

$$R_i^\star = \frac{1}{n-m+i}\left[\ln\left\{\nu\gamma_0^{n-m+i}(n-m+i-1)!\right\}\right]_+ \quad (52)$$

where the operator $[\cdot]_+ = \max[0, \cdot]$ accounts for the case $R_i^\star = 0$. While this approximation becomes less accurate for $R_i \leq 1$, at least one of the $R_i^\star$ satisfies $R_i^\star > R$, and hence the approximation is accurate for at least one of the $R_i$ if the total rate is not too small, $mR \geq 1$. The approximation error in smaller $R_i^\star$ does not have as much impact on the total outage probability so that the over accuracy is good. The Lagrange multiplier $\nu$ is found by substituting $R_i^\star$ from (52) into the total rate constraint $\sum R_i = mR$:

$$mR = \sum_{i=m-m_A+1}^{m} \frac{\ln \nu + \ln\left[\gamma_0^{n-m+i}(n-m+i-1)!\right]}{n-m+i}$$
$$\Rightarrow \nu = \gamma_0^{\frac{-m_A}{b}} e^{\frac{mR-a}{b}} \qquad (53)$$

### C. Proof of Lemma 5 and Theorem 3

*Proof of Lemma 5 (convexity):* Consider the following function: $f(R, \alpha) = (e^R - 1)/\alpha$. Its Hessian

$$\triangle f = \begin{bmatrix} e^R/\alpha & -e^R/\alpha^2 \\ -e^R/\alpha^2 & 2(e^R-1)/\alpha^3 \end{bmatrix}$$

is positive definite at the moderate to high rate regime, $R > \ln 2$, since the diagonal entries are positive and the determinant is also positive, $\det \triangle f = e^R(e^R - 2)/\alpha^4 > 0$. Hence, $f(R, \alpha)$ is jointly convex in $(R, \alpha)$, $P_i$ is convex by the composition rule [22], and $P_{out}$ is convex as a sum of convex functions. Since the constraints are also convex, the APRA problem is convex. ∎

*Proof of Theorem 3 (closed form solution):* The Lagrangian for this problem is

$$L = \sum_i^m P_i - \nu_R \left(\sum_i^m R_i - mR\right) - \nu_\alpha \left(\sum_i^m \alpha_i - m\right)$$
$$- \sum_i^m \lambda_{R_i} R_i - \sum_i^m \lambda_{\alpha_i} \alpha_i$$

and the KKT conditions are given by

$$R_i^\star \geq 0, \ \sum_i R_i^\star = mR, \ \lambda_{R_i} \geq 0, \ \lambda_{R_i} R_i^\star = 0$$
$$\alpha_i^\star \geq 0, \ \sum_i \alpha_i^\star = m, \ \lambda_{\alpha_i} \geq 0, \ \lambda_{\alpha_i} \alpha_i^\star = 0$$
$$P_i'(R_i^\star) - \lambda_{R_i} - \nu_R = 0, \ -\frac{P_i'(R_i^\star)}{\alpha_i^\star} - \lambda_{\alpha_i} - \nu_\alpha = 0, \quad (54)$$

Here, $P_i'$ denotes derivative of $P_i$ with respect to $R_i$, and we have exploited the fact that

$$\frac{dP_i}{d\alpha_i} = -\frac{e^{R_i} - 1}{e^{R_i}} \frac{1}{\alpha_i} \frac{dP_i}{dR_i} \approx -\frac{1}{\alpha_i} \frac{dP_i}{dR_i}$$

when $1 < R_i < \ln \gamma_0$, as easy to verify from (5). As before, $\lambda_{R_i}$ and $\lambda_{\alpha_i}$ are slack variables and can be eliminated from (54):

$$\begin{cases} P_i'(R_i^\star) - \nu_R \geq 0 \\ -\frac{P_i'(R_i^\star)}{\alpha_i^\star} - \nu_\alpha \geq 0 \end{cases} \qquad (55)$$

Expanding $\lambda_{R_i} R_i^\star = 0$, $\lambda_{\alpha_i} \alpha_i^\star = 0$ while using (54), one obtains:

$$\begin{cases} (P_i'(R_i^\star) - \nu_R) R_i^\star = 0 \\ \left(-\frac{P_i'(R_i^\star)}{\alpha_i^\star} - \nu_\alpha\right) \alpha_i^\star = 0 \end{cases} \qquad (56)$$

If $i$-th transmitter is active, $R_i^\star > 0$, $\nu_R = P_i'(R_i^\star)$ and $\alpha_i^\star = -P_i'(R_i^\star)/\nu_\alpha^\star = -\nu_R/\nu_\alpha$ does not depend on $i$. We conclude that the uniform power allocation among the active streams is optimal. According to the total power constraint, the amount of power allocated to each of the active streams is equal to $m/m_A$, where $m_A$ is the number of active streams. Hence the optimal rates are given by the same expressions as in the ARA in (21), with $\gamma_0$ changed to $\frac{m}{m_A}\gamma_0$. ∎

### D. Proof of Theorem 4

A formal proof can be constructed using Lagrange duality theory [22] (using the fact that the duality gap is zero in this case). A more simple and straightforward proof is by contradiction. Assume that $\boldsymbol{\alpha}^*$ solves (8) (the solution is unique since the problem is strictly convex) and that $\boldsymbol{\alpha}^0$ solves (45) with $\epsilon = P_{out}(\boldsymbol{\alpha}^*)$. Since $P_{out}$ is strictly decreasing in $\alpha_i$, the solution in (45) is on the boundary $P_{out} = \epsilon$ so that $\sum_{i=0}^m \alpha_i^0 \leq \sum_{i=0}^m \alpha_i^* = m$. Define

$$\beta = \frac{m}{\sum_{i=0}^m \alpha_i^0} \geq 1 \qquad (57)$$

so that

$$P_{out}(\beta \boldsymbol{\alpha}^0) \leq \epsilon = P_{out}(\boldsymbol{\alpha}^*) \qquad (58)$$

If strict inequality holds in (58), then there is a contradiction since $\boldsymbol{\alpha}^*$ is the minimizer of $P_{out}$. If equality holds in (58) and $\beta \boldsymbol{\alpha}^0 \neq \boldsymbol{\alpha}^*$, then there is a contradiction since $\boldsymbol{\alpha}^*$ is the unique minimizer. The only remaining possibility is $\beta \boldsymbol{\alpha}^0 = \boldsymbol{\alpha}^*$ so that $\boldsymbol{\alpha}^0 = \boldsymbol{\alpha}^*$.

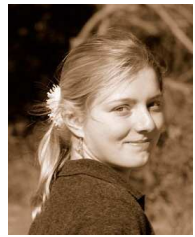

**Victoria Kostina** received a Bachelor's degree with honors in applied mathematics and physics from the Moscow Institute of Physics and Technology, Russia, in 2004, and a Master's degree in electrical engineering from the University of Ottawa, Canada, in 2006. She is currently working towards her Ph.D. at Princeton University, USA. Her research interests are in information theory, coding and wireless communications.

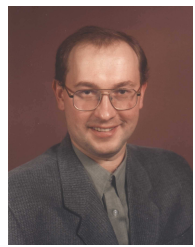

**Sergey Loyka** (M'96–SM'04) was born in Minsk, Belarus. He received the Ph.D. degree in Radio Engineering from the Belorussian State University of Informatics and Radioelectronics (BSUIR), Minsk, Belarus in 1995 and the M.S. degree with honors from Minsk Radioengineering Institute, Minsk, Belarus in 1992. Since 2001 he has been a faculty member at the School of Information Technology and Engineering, University of Ottawa, Canada. Prior to that, he was a research fellow in the Laboratory of Communications and Integrated Microelectronics (LACIME) of Ecole de Technologie Superieure, Montreal, Canada; a senior scientist at the Electromagnetic Compatibility Laboratory of BSUIR, Belarus; an invited scientist at the Laboratory of Electromagnetism and Acoustic (LEMA), Swiss Federal Institute of Technology, Lausanne, Switzerland. His research areas include wireless communications and networks, MIMO systems and smart antennas, RF system modeling and simulation, and electromagnetic compatibility, in which he has published extensively. Dr. Loyka is a technical program committee member of several IEEE conferences and a reviewer for numerous IEEE periodicals and conferences. He received a number of awards from the URSI, the IEEE, the Swiss, Belarus and former USSR governments, and the Soros Foundation.